\documentclass[journal,twocolumn,10pt]{IEEEtran}
\usepackage{graphicx}
\usepackage{epstopdf}
\usepackage{amsmath}
\usepackage{amsfonts}
\usepackage{amssymb}
\usepackage{algorithm}
\usepackage{algpseudocode}
\usepackage{authblk}
\usepackage{xcolor}
\usepackage{hyperref}

\begin{document}
\title{Localisation, Communication and Networking with VLC: Challenges and Opportunities}

\author[1]{Rong Zhang}
\affil[1]{School of Electronics and Computer Science, University of Southampton, UK}

\markboth{}{}

\maketitle

\begin{abstract}
The forthcoming Fifth Generation (5G) era raises the expectation for ubiquitous wireless connectivity to enhance human experiences in information and knowledge sharing as well as in entertainment and social interactions. The promising Visible Light Communications (VLC) lies in the intersection field of optical and wireless communications, where substantial amount of new knowledge has been generated by multi-faceted investigations ranging from the understanding of optical communications and signal processing techniques to the development of disruptive networking solutions and to the exploitation of joint localisation and communications. Building on these new understandings and exciting developments, this paper provides an overview on the three inter-linked research strands of VLC, namely localisation, communications and networking. Advanced recent research activities are comprehensively reviewed and intriguing future research directions are actively discussed, along with the identifications of a range of challenges, both for enhancing the established applications and for stimulating the emerging applications.
\end{abstract}


\section{Introduction}
We are at the dawn of an era in information and communications technology with unprecedented demand for digitalised everything, for connected everything and for automated everything~\cite{book}. The next decade will witness a range of disruptive changes in wireless technologies, embracing all aspects of cross-disciplinary innovations~\cite{6191306}. Fundamental challenges arise when we have reached the limit of the conventional Radio Frequency (RF) based wireless technologies, which are increasingly less capable of meeting the escalating traffic-demands and of satisfying the emerging use-cases. Especially, there have been substantial research efforts dedicated to the high carrier frequencies, including the millimetre wave~\cite{6515173} and the visible light spectrum~\cite{6963803} in the forthcoming Fifth Generation (5G) wireless networks landscape. In this paper, we provide an overview on the fast growing technology of Visible Light Communications (VLC), which lies at the cross-section of optical and wireless communications, and focuses on the human perceivable part of the electromagnetic spectrum, corresponding to wavelengths from 380nm to 780nm. 

The use of the visible light spectrum for wireless communications has gained great interests. This is because the visible light spectrum is licence-free, has a vast bandwidth and does not interfere with the RF band. Historically, in the late 19th-century, Alexander Graham Bell invented the photo-phone by transmitting voice signals over modulated sunlight~\cite{5300478}. Almost a century later, artificial light generated by fluorescent lamps was also successfully demonstrated for supporting low data-rate communications~\cite{673734}. It is these exciting experiments that inspired the modern VLC using Light Emitting Diodes (LEDs). In addition to facilitating communications, being a modern illumination technology as their main function, LEDs have been increasingly dominating over the traditional incandescent lamps and fluorescent lamps, owing to their higher energy-efficiency, colour-rendering capability and longevity~\cite{5456162}. Hence, the potential for VLC is further supported by the anticipated presence of a ubiquitous and efficient LED lighting infrastructure. 

The pioneering implementation of VLC using LEDs for the dual purpose of indoor illumination and communications was carried out by the Nakagawa laboratory in the early 2000s~\cite{1277847}. Subsequently, tremendous research efforts have been invested in improving the link-level performance between a single LEDs array and a single receiver, focusing on both the LED components and on the VLC transceivers, where ambitious multi Gbps targets have been achieved. These exciting link-level achievements set the basis for broadening the scope of VLC research beyond point-to-point applications~\cite{6852089}, with the focus on VLC aided networking, where various promising protocols, architectures and cross-layer solutions have been proposed. Furthermore, LEDs based localisation has also attracted dedicated research interests, where recent advances have demonstrated sub-centimetre accuracy and 3D positioning capability~\cite{7096289}. In addition to the main thrust research in localisation, communications and networking, VLC can also provide innovative solutions for a number of use-cases~\cite{7239528}, including vehicular, Device-to-Device (D2D) and underwater applications, just to name a few. 

Along with the above technical advances, there have been significantly increased activities in the VLC domain. Large scale research programmes were launched, bringing together research-led universities and industries, as exemplified by the Visible Light Communications Consortium (VLCC), the EU-FP7 project OMEGA, the consortium on ultra parallel VLC. Most recently, VLC was also included in the scope for networking research beyond 5G under the framework of EU H2020. Dedicated research centres have been also established, such as the Smart Lighting Engineering Research Center in US, the Li-Fi Research and Development Centre in UK and the Optical Wireless Communication and Network Centre in China, etc. Furthermore, in recent years, ComSoc has published three special issues on VLC in the Communications Magazine~\cite{6685753,6852084} and in the Wireless Communications Magazine~\cite{Wcom}. Moreover, three consecutive international workshops on VLC have also been sponsored by ComSoc along with ICC'15, ICC'16 and ICC'17. Meanwhile, GlobeCom has continuously offered annual workshops on optical wireless since 2010. These investments and efforts have all contributed to the growing success of the subject.

\begin{figure}[t]
	\centering
	\includegraphics[trim={0.25cm 5.75cm 0.25cm 0.5cm},clip,width=\linewidth]{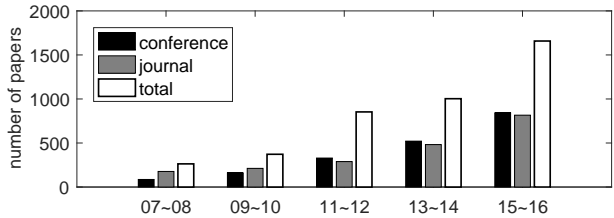} 
	\caption{Biannual statistics on the number of papers published in IEEE and OSA in the topic of visible light communications and positioning. These data are gathered by searching for the keywords (Visible Light Communications, VLC, Visible light Positioning, VLP).}
	\label{fig_paper}
\end{figure}

Hence, we see a strong momentum on the research of VLC, as evidenced by the publication statistics in Fig~\ref{fig_paper}, that motivates this special issue and our paper is organised as follows. We will first introduce some basics in Section~\ref{sec_basic}. We will then review the achievements of three main research strands of VLC, namely localisation in Section~\ref{sec_localisation}, communications in Section~\ref{sec_comms} and networking in Section~\ref{sec_networking}. After that, we will discuss about the challenges of VLC in Section~\ref{sec_challange} and paint our vision on the future of VLC in Section~\ref{sec_vision}. Finally, we will conclude our discourse in Section~\ref{sec_con}.

\section{Main Research Strands}
\label{sec_essential}
\subsection{Basics}
\label{sec_basic}
The modern VLC relies on LEDs for the dual purpose of illumination and communications.  LEDs are solid-state semiconductor devices~\cite{5456162,4982622}, which are capable of converting electrical energy to incoherent light through the process of \textit{electro-luminescence}. When using LEDs for VLC, data can be readily modulated by electronically flickering the intensity of light at a rate fast enough to be imperceivable by human eyes. At the receiver, data can be directly detected using inexpensive photo-detectors or imaging sensors. This transceiver structure may be referred to as the Intensity Modulation (IM) and Direct Detection (DD), which is much simpler than the conventional coherent RF receivers relying on complicated and energy-consuming heterodyne structure. Although there exist implementations of Laser Diodes (LDs) for VLC~\cite{Lee:15,7457221} that can provide an even higher data-rate, our following discussions will be based on the more popular LEDs.  

The primary function of LEDs is for illumination, hence the performance of VLC is naturally dependent on the specific characteristics of LEDs~\cite{7107983}. White light is the most widely used color for many applications, where there are three different methods to produce while light LEDs. The most common method is the use of blue light LEDs with yellow phosphor coating layer~\cite{999186}. This is a low-cost method, but the coating layer limits the modulation bandwidth to only tens of MHz. Early experiments have demonstrated tens of Mbps data-rate~\cite{4547916}, while hundreds of Mbps data-rate may also be achieved with advanced processing~\cite{4926203} and adaptive receiver~\cite{6646247}. The second method to generate white light is to carefully mix the color of Red, Green and Blue (RGB) emitted from the RGB LEDs~\cite{999188}. This method is more expensive, but it offers much wider modulation bandwidth upto hundreds of MHz, supporting multi Gbps data-rate~\cite{Cossu:12}. The third method is resulted from the recent advances on micro-LED array~\cite{6072221}, which produces white light through wavelength conversion. This method is capable of facilitating parallel communications with even higher data-rate~\cite{7492305}. 

When the receiving side is considered, there are generally two types of detectors that are often employed, namely the photo-detectors and the imaging sensors~\cite{5629350}. In particular, the Positive-Intrinsic-Negative (PIN) photo-detectors are most widely considered in current studies, owing to its low cost to implement and high sensitivity to strong incident light. By contrast, in the scenario of weak incident light, avalanche photo-detectors are often more preferred. On the other hand, imaging sensors are constituted by an array of photo-detectors, which inherently support the use of pixelated reception owing to its desirable capability in better separating the optical channels~\cite{1668130}. Another advantage of imaging sensors is that they can be readily integrated with camera lens in smart hand-held devices~\cite{7516566}. Since the frame-rate of the commonly used camera lens is low, this type of receiver is particularly suitable for applications requiring low date-rate~\cite{7265146}, such as the project of CeilingCast~\cite{7524511}. Sometimes, focusing lens may also be used on top of photo-detectors~\cite{6497451} or imaging sensors~\cite{6891135} to enhance the reception by collecting more light energy from diffuse links with increased Field of View (FoV). In addition to the above receivers, other interesting developments include prism-array receiver~\cite{7106458}, solar-cell receiver~\cite{7355280}, rotational receiver~\cite{7439739} and aperture-based receiver~\cite{7786897}, etc.

Since the surface diameter of typical photo-detectors is several thousand times of the light wavelength, the fast fading effects are averaged out at the receiver. Hence, when the indoor propagation is considered, the Line of Sight (LoS) path-loss effect is typically observed, plus the reflections from diffuse links~\cite{6781654}. In most of the studies, (dispersive) channel modelling is carried out by assuming Lambertian type emitters and for simplicity, we adopt the well-established Infra-Red (IR) channel modelling~\cite{5682214}, with the aid of the convolution-based simulation method to properly characterize the Power Delay Profile (PDP) of the propagation channels~\cite{7339420}. This method comes at the expense of complexity, hence more efficient computational methods were proposed~\cite{7029002}. Furthermore, there are three typical sources of noise, namely the ambient noise, short noise and thermal noise. In fact, they are mutually dependent, hence it is important to carry out system design by taking into account this property~\cite{7723939}. 

\subsection{Localisation}
\label{sec_localisation}
The widely used Global Navigation Satellite System (GNSS) technology provides a coarse localisation capability that works just fine for the majority of outdoor applications~\cite{7445788}. However, its use for indoor localisation fails, since the satellite signal has a poor indoor penetration and the multi-path reflections are highly complex in indoor environment. Various alternatives were proposed for indoor localisation services~\cite{4343996}, mainly based on RF techniques such as Wireless Fidelity (Wi-Fi) or Ultra-Wide Band (UWB). However, the associated cost is often too high to reach the desired accuracy. Hence, LEDs based localisation becomes an interesting option to meet the above demand~\cite{6685759}. This is because the LEDs can be readily integrated in the existed lightening infrastructure and several nearby LEDs may provide joint localisation to achieve a very high level of accuracy. Furthermore, the visible light spectrum can be used in RF sensitive environment and the imaging sensors built-in the smart hand-held devices constitute a convenient receiver for LEDs based localisation, or inexpensive PDs may be purposely installed. In the future 5G systems, both manufacturing and service robots will become dominant Machine Type Communication (MTC) devices. In homes, factories and warehouses, light is useful for accurate position control of robots~\cite{7752576}, again thanks to its linear propagation property and short wavelength. Explicitly, there exist several approaches for LEDs based localisation, including the proximity based approach, the triangulation based approach and the fingerprinting based approach, where each of them will be elaborated in detail as follows. 

The simplest way to realise LEDs based localisation is the \textit{proximity} based approach, where the position of the object is roughly classified to its nearest LED transmitter. This procedure can be conveniently integrated in the cell search stage, where no complicated algorithms are imposed~\cite{7469371}. Hence, the achievable localisation accuracy is very coarse, in terms of decimetres. When further exploiting the geometric properties of multiple LEDs, the \textit{triangulation} based approach is capable of determining the absolute position of the object and reaching a localisation accuracy in terms of centimetres. This approach exploits the measured distance (angles) between the multiple LEDs and the localisation object, where the measurement could be Time of Arrival (ToA)~\cite{6600748}, Time Difference of Arrival (TDoA)~\cite{6131130}, Received Signal Strength (RSS)~\cite{6880333} and Angle of Arrival (AoA)~\cite{6698953}. There are several challenges associated with these measurements that may deteriorate the localisation accuracy. For example, both ToA and TDoA require strict timing synchronisation, RSS requires detailed knowledge of radiation pattern and AoA requires careful calibration of light directionality. Hence, localisation based on a combination of different measurements is a good practice~\cite{6823667,7339418}, particularly for 3D localisation~\cite{7247633}. Finally, the \textit{fingerprinting} based approach determines the position of the object by comparing on-line measured data with pre-stored off-line measured data~\cite{7468514}. Depending on the richness of the pre-stored data set, this approach may provide a higher level of localisation accuracy, but at the cost of an increased complexity. Nevertheless, this approach has a poor scalability to use owing to its scene-dependence. 

Unique approach exists for LEDs based localisation using camera sensors available in smart hand-held devices~\cite{7564490}. Thanks to the high density of pixels, camera sensors can extract detailed spatial information in order to determine the position of the object with high localisation accuracy by using vision analysis and image processing. In addition to only use camera sensors, accelerometer build-in the smart hand-held devices can also be exploited together to achieve 3D localisation~\cite{6868970,s16060783}. To conclude, despite the existence of various LEDs based localisation approaches and their achievements, LoS blockage constitutes the single dominant problem, which is lacking of considerations in the current research. Hence, the recent study on the effect of multi-path reflections on positioning accuracy becomes highly practical~\cite{7431937}. More importantly, when localisation is considered, it should be included in the entire design chain for achieving a better integrity~\cite{7045480}. We believe that with further scientific advances, powerful and robust solutions for LEDs based localisation will be developed to meet diverse requirements. 

\subsection{Communications}
\label{sec_comms}
As discussed above, VLC is capable of realising the dual purpose of illumination and communications. Amongst the cascaded physical layer components of VLC, we elaborate on those that require contrived design and dedicated discussions, namely optical domain modulation and Multiple Input Multiple Output (MIMO). Generally, there are two types of modulation techniques that are commonly employed for VLC, namely the \textit{single-carrier} modulation and \textit{multi-carrier} modulation. To elaborate on the former, On-Off Keying (OOK) constitutes the simplest technique for VLC modulation. Hence, it is the most commonly studied and experimented technique together with various different types of LEDs. Despite its simplicity, multi Gbps data-rate has been reported in recent experiment~\cite{7505962}. Developed from the plain OOK, there is a family of pulse-based modulations, such as Pulse Width Modulation (PWM)~\cite{6530832} and Pulse Position Modulation (PPM)~\cite{7284753}. These modulations belong to the M-ary orthogonal signalling, which is particularly suitable for IM. Unique to VLC, Color Shift Keying (CSK) has been introduced in the VLC standardisation 802.15.7 for simultaneously achieving a higher data-rate and a better dimming support~\cite{6780585,7008542}. Specifically, CSK modulates the signal based on the intensity of the RGB color, relying on the employment of RGB LEDs. Other interesting schemes exploiting the color property of VLC may be found in~\cite{7054450,6998813}. It is worth noting that as a modulation technique, CSK is different from the concept of Wavelength Division Multiplexing (WDM)~\cite{7454689}. This is because in WDM system, additional modulation may be designed together with each of the multiplexing layers.

When considering the multi-carrier modulation family, the celebrated Optical Orthogonal Frequency Division Multiplexing (OOFDM) schemes are often employed. This is because OOFDM scheme allows parallel data transmissions, and it is also capable of combating the detrimental channel dispersion without complex time domain equalisations. Different from the conventional RF based OFDM schemes, transmitted signals of OOFDM need to be real-valued and positive in order to facilitate IM. There are a family of OOFDM schemes, typically including the Asymmetrically Clipped OOFDM (ACO-OFDM) scheme and the DC-biased OOFDM (DCO-OFDM) scheme~\cite{4785281}. In both schemes, the property of Hermitian symmetry is exploited for rendering the complex-valued signal to be real-valued, at the cost of halving the bandwidth efficiency. In order to maintain positivity of the transmitted signal, ACO-OFDM scheme resorts to use only the odd sub-carriers, while DCO-OFDM scheme resorts to apply sufficient DC bias. Hence, the former approach is more power efficient, while the latter approach is more bandwidth efficient~\cite{5967993}. Moreover, there exist many other interesting realisations of OOFDM, such as the flip OOFDM~\cite{6291705}, unipolar OOFDM~\cite{7106494}, multi-layer OOFDM~\cite{6685605}, hybrid OOFDM~\cite{6821328}, and DC-informative OOFDM~\cite{7096285}. In general, OOFDM schemes suffer from the classic problem of high Peak to Average Power Ratio (PAPR)~\cite{Wang:16} and their performance is also limited by the clipping distortion~\cite{6162931} and the LEDs' non-linearity~\cite{6395783}, where many countermeasures have thus been proposed~\cite{6837447,7096283,7208804}. Despite all these challenges, OOFDM schemes have attracted great attentions owing to their high data-rate potential, robustness to channel dispersion and flexibility in resource allocation~\cite{6965617}. 

It is important that the above modulation techniques should be co-designed with LEDs' illumination requirements to avoid flickering and support dimming~\cite{6685755}. Flickering refers to the undesirable effect of human perceivable brightness fluctuation, or in other words light intensity changes. This is relatively easy to cope with by using Run Length Limited (RLL) coding in order to balance the resultant zeros and ones~\cite{7564553,7397916}. On the other hand, modern LEDs have now been capable of supporting arbitrary levels of dimming for energy saving and environmental-friendly illumination. Hence, a more intriguing aspect is to jointly design modulation and dimming~\cite{5773477}. In general, there are two approaches to support dimming with modulation, namely to control either the light intensity or the signal duty cycle, where the former is easier to implement and the latter achieves higher precision~\cite{7096282}. As an example, in OOK, its on or off levels can be defined to support dimming or one may introduce compensation period along with the signalling period without resorting to intensity modification. Amongst others, pulse-based modulations show great flexibility in support dimming, such as the variable PPM scheme proposed in the 802.15.7 standard~\cite{6163585}. On the other hand, dimming support in multi-carrier modulation requires further investigations, where recent research has been dedicated to this direction~\cite{7046383,7484227}. In addition to modulation, channel coding schemes can also be co-designed with dimming support in mind, as demonstrated in various research efforts, including turbo codes~\cite{6198865}, Reed-Muller codes~\cite{6338266}, adaptive codes~\cite{6937164} and concatenated codes~\cite{7518602}. 

Similar to the conventional RF based wireless communications systems, MIMO in VLC is also capable of providing promising throughput enhancements~\cite{6376096}. It is known that the full potential of MIMO can only be achieved in a fully scattered environment. However, in VLC, the particular challenge is that the optical MIMO channels are often highly correlated and the resultant MIMO channel matrix appears to be rank deficient~\cite{5342325}. To create full rank MIMO channel matrix, it is crucial to maintain clear separation of the transmitters at the (non)-imaging receiver by careful calibration. In addition to support full rank MIMO channel matrix, robustness to receiver tilts and blockage is also desirable, where it has been shown that specifically designed receivers to harness angle diversity constitutes a good design practice~\cite{7284696,7109107}. Most recently, the research on MIMO in VLC under diffuse channel is also emerging, leading to substantial practical insights~\cite{7335563}. As far as the MIMO functions are considered, MIMO in VLC can achieve diversity gain by using space time coding~\cite{7000619}, multiplexing gain by using parallel transmission~\cite{6476612} and beamforming gain by using electronic beam steering~\cite{7494883}. Multiple transmit luminaries can also be used to improve the security of VLC transmission by making the VLC signal difficult to intercept by an eavesdropper. Several linear beam-forming schemes for active and passive eavesdroppers have recently been presented in~\cite{7106457,7555388,7556955}. Being an important generalisation, Multiple Input Single Output (MISO) transmission and in particular, Multi-User MISO (MU-MISO) transmission have attracted substantial research interests~\cite{7124415}. This is because the MU-MISO scheme provides beneficial multi-user diversity gain without incurring rank deficient MIMO channel matrix. However, in this multi-user scenario, challenges arise when performing inter-user interference cancellation, where (non)-linear transmit pre-coding scheme is required. 

\subsection{Networking}
\label{sec_networking}
The above mentioned advances in physical layer research of VLC have lead to the development of VLC aided networking~\cite{7360112}. Being next to the physical layer, there are several different candidates for the Medium Access Control (MAC) layer, including both the \textit{multiple access} and \textit{random access} schemes. Let us now elaborate on the multiple access scheme first, the most straightforward arrangement is the Time Division Multiple Access (TDMA) scheme~\cite{7107881,7707381}, where users simply access the network in different time slots. When multi-carrier modulation is employed, the Orthogonal Frequency Division Multiple Access (OFDMA) scheme allows users to be allocated different Time and Frequency (TF) resource blocks~\cite{6747983}. When compared to TDMA scheme, OFDMA scheme provides a higher flexibility in terms of resource allocation and user scheduling, at a modestly increased complexity. In addition to the above orthogonal multiple access schemes, in Non-orthogonal Multiple Access (NOMA) scheme~\cite{7275086,7572968}, two or more users may be multiplexed in power domain in addition to the conventional orthogonal TF domain. At the receiver side, onion-stripping type interference cancellation is required to separate the users from power-domain non-orthogonal multiplexing~\cite{5522468}. Other than relying on the power-domain, spatial domain could also be exploited at the transmitter to realise NOMA~\cite{7494887}. Differently, (multi-carrier) Optical Code Division Multiple Access (OCDMA) scheme relies on assigning each user a unique and specific optical code~\cite{7174962,7116868}. Finally, when random access is considered, the classic Carrier Sense Multiple Access (CSMA) scheme remains highly attractive. Importantly, early implementations have already shown its successful usage~\cite{6852087} and in slotted random access, both contention access periods and free periods are included, where the latter ensures guaranteed time slots for resource limited applications. 

The most straightforward way of constructing an indoor VLC cell is to simply consider each Access Point (AP) function as an individual cell and to adopt the unity frequency reuse across all cells. This construction would result in the highest spatial reuse but it tends to suffer from the typical problem of Inter-Cell Interference (ICI) amongst neighbouring cells. Following the traditional cellular design principle~\cite{5641646}, different (fractional) frequency reuse patterns could be used to mitigate ICI, at the cost of a reduced bandwidth efficiency. An effective method of improving the efficiency, whilst mitigating the detrimental ICI is to employ cell merging~\cite{7096279}, where a group of neighbouring APs jointly form an enlarged VLC cluster. In this way, the previously ICI-contaminated area becomes the cluster-centre of the newly formed cluster. Multiple users can be served simultaneously by using sophisticated Vectored Transmission (VT) techniques. The underlying principle is to totally eliminate the ICI at the multiple transmitters side by using transmit pre-coding, so that the multiple users receive mutually interference-free signals~\cite{5640682}. However, this technique requires that both Channel State Information (CSI) and the users' data have to be shared amongst multiple APs. All of the above cell formations follow the conventional \textit{cell-centric} design approach, which is based on defining a cell constituted by a fixed set of one or more APs and then associating the users with it. By contrast, the newly proposed \textit{user-centric} design approach relies on the dynamic association between APs and users~\cite{7217841}. More explicitly, by taking into account the users' geo-locations, the new user-centric design flow is based on grouping the users together and then associating the APs with them, leading to amorphous cells~\cite{7437435}, which is capable of supporting video service on the move~\cite{7802594}. Finally, an intriguing piece of research emerges to consider the optimal placement of LEDs~\cite{7328268}, resulting into rich implications on throughput, delay and mobility.

Holistically, owing to the existence of lightening infrastructure, Power-Line Communications (PLC) constitute a convenient back-haul for VLC as indoor access technology~\cite{6525866,7052407}. PLC can reach luminaries that serve as VLC transmitters to supply the data streams as well as to coordinate transmission between multiple VLC transmitters to support multi-user broadcasting~\cite{7147818}. The VLC transceivers can be considered as relays that can operate in a full-duplex model and different relaying paradigms such as amplify-and-forward and decode-and-forward are possible~\cite{6525866}. From a networking perspective, VLC can be considered as a new member in the small-cell family of the Heterogeneous Networks (HetNet) landscape for complementing the over-loaded Radio Access Technology (RAT). Indeed, the interplay between VLC and RAT system has been an active area of research~\cite{6637084,7293077}, where there are two different interplay scenarios that may be envisioned, namely \textit{single-homing} and \textit{multi-homing}. In the single-homing scenario, only one access system is allowed to maintain its association at any instant. In this scenario, dynamic Load Balancing (LB) will prevent traffic congestion caused by blockage or mobility through diverting the traffic flow appropriately~\cite{7056535}. To better exploit the access system's diversity potentials, in the multi-homing scenario, each user maintains multiple associations at the same time by using the Multipath Transport Control Protocol (TCP) to connect multiple interfaces~\cite{6933944}. In either scenario, robust vertical handover has to be properly designed to mitigate any ping-pong effect, where the load-ware mobility management appears to be a promising solution~\cite{7274270}. Different from using higher layer converging approaches, seamless rate adaptation between VLC and RAT systems may also be achieved by network coding and rate-less coding~\cite{7096290}. To sum up, in the light of the information and communications technology convergence, the above mentioned network layer functions should be soft-ware defined to maximise its full potential~\cite{7473831,7503119}. 

\section{Challenges and Opportunities}
\label{sec_future}
\subsection{Challenges}
\label{sec_challange}
\subsubsection{Channel Modelling}
Most of the current channel modelling in VLC was directly adapted from the IR communications. However, it would be ideal to develop specific VLC channel modelling, corresponding to different types of LEDs for the ease of cross-calibration. In particular, with regards to shadowing, there is a lack of both empirical and statistical modelling. In most of the studies, the shadowing effect was often assumed to follow the over-simplified Bernoulli distribution. However, given the sensitivity of VLC over shadowing in all aspects of localisation, communications and networking, a well-calibrated model is indeed of critical importance. Also importantly, VLC channel modelling for vehicular applications is still in its infancy, which requires dedicated efforts.

\subsubsection{Interference Mitigation and Management}
VLC relies on visible light spectrum which overlaps with solar radiation and indoor/outdoor lighting and display. A VLC system is inevitably interfered by those sources, from day and night time. Therefore, effective techniques are needed to mitigate not only inter-system interference from neighbouring LEDs, but also external light interference. Considering those interference sources typically emit a large dynamic range of light intensities, a robust and sensitive detector needs to respond reliably to transmitted signals while suppressing interference to certain extent. It is necessary but very challenging to develop advanced methods and algorithms to recover useful signals, weak or strong, from noisy signals overwhelmed by interference, weak or strong. For example, normal operation of a vehicular receiver or an indoor receiver under direct sunlight through a window at noon is extremely difficult.

\subsubsection{Channel Feedback}
Several contributions make a massive use of channel state information. While this aspect is not critical in approaches that move all the processing toward the receiver since estimation techniques are not so difficult to implement both with training-based or (semi)-blind, when the communications requires a feedback link, as for pre-equalization, bit loading, retransmission schemes, the feedback channel performance becomes an issue. The problem of how a feedback channel is used is a very challenging task either if it is an optical one and if it is based on RF. In fact, important aspects for example quantisation and error control must be properly taken into account, in addition to delay and jitter.

\subsubsection{Access Techniques}
A high percentage of access techniques proposed in the literature and described above start from the tentative of adapting access mechanisms and procedures already used in the RF for implementing them into the VLC context. However, mechanisms that take into account the fact that VLC receivers usually have one or more photo-detectors that are not omnidirectional as well as for emerging VLC applications such as vehicular and camera communications, should be investigated. Access techniques become more complicated, when a user is \textit{covered} by some LEDs in indoor environment, due to the tilt of the device, the signal can be received from another set of LEDs. Hence, access techniques are strictly related also to the localisation techniques.

\subsubsection{Mobility Management}
In many VLC applications, VLC terminals or transceivers are mobile, such as hand-held devices, vehicles, and robots. Maintaining a quality communications link in a point-to-point case and network connectivity in a multiple-user case are important to avoid communications losses. Protocols for mobile optical wireless communication networks are in urgent need, for dynamic system resource adaptation to communications environments, easy node access and drop-off, smooth handover from one AP to another and from one network to another. Meanwhile, mobility prediction and network topology modelling are under-explored, and these topics open new room for further investigation.

\subsubsection{Integrated Smart Lighting}
 Since VLC is not a paradigm separated from illumination, the ultimate future challenge can be a full integration that takes into account not only communications performance related to the single link under illumination constraints but also a holistic optimisation regarding both the networking (system performance and seamless handover) and the perceived illumination (a good level of light all over the room). This is an issue especially in large rooms such as museums, where multimedia content may be based on the position of the user in the information centric system. Hence, the aim is to integrate lighting, communications, networking and positioning in a single homogeneous framework.

\subsection{Opportunities}
\label{sec_vision}
\subsubsection{5G-Home}
The concept of 5G-Home is based on the unified heterogeneous access-home network with wired and (optical) wireless connectivity, which is capable of creating tens of millions of additional 5G-Home sites. Technically, to succeed in a timely and affordable manner, both the existing copper or cable based fixed access and in-home wiring technique need to be exploited in addition to fibre~\cite{7467432,7600493}. Deep in the home, the number of wireless APs and terminals will be going up corresponding to the increasing bandwidth, delay and coverage requirements. At least one AP per room may become the norm given the higher carrier frequencies that will be used for 5G, including Wi-Fi standards 802.11ac/ad and VLC aided networking~\cite{6059258,7402263}. Hence, we envision beneficial convergence between fixed access network infrastructure and in-home (optical) wireless networks, in order to eliminate the boundary between these two domains~\cite{7470948}.

\subsubsection{5G-Vehicle}
Future 5G systems will embody a number of new applications which span across vast areas beyond enhanced mobile broadband services, such as media distribution, Smart Cities, and Internet of Things (IoT). In Smart Cities, light-enabled vehicular communications networks utilize a large number of densely distributed street lamp poles as APs, while vehicular lights, roadside street lights, pedestrian signage and traffic lights can all act as ubiquitous transmitters to meet the need of massive connectivity and high throughput~\cite{Cui:12,7317859}. Equipped with image sensor or other types of receivers, these nodes in the immediate vicinity of vehicles provide ultra-reliable and low latency (mission critical) communications and control links~\cite{6852088,7454688}, which will serve well intelligent transportation, collision avoidance, autonomous driving and telematics. However, challenges will arise when experiencing strong ambient noise, while uneven speed and diverting route may be also difficult to handle~\cite{7105832,7406675}. Nevertheless, vehicles become mobile offices and entertaining homes, enjoying amount of exterior resources.

\subsubsection{Underwater Communications}
In addition to ground-based indoor and outdoor applications, VLC also paves the way for outreach applications deep into water. Currently, underwater acoustics is the major wireless technology for underwater object detection and communications. Due to the very low data rate allowed by acoustic communications in underwater environment and also the poor performance from the propagation point of view in the RF bands, light sources offer a unique opportunity for short-range high speed communications potentially at low cost and low power~\cite{6876673,6986327,Wang2:16,Li:16}. Data rate of hundreds to thousands of Mbps is possible, and the communications distance is extendible to hundreds of meters in clear water. The VLC technology will undoubtedly play a critical role in marine resource exploration, water sensing, monitoring and tracking of marine organism, and saving endangered species.

\subsubsection{Emerging Applications}
LEDs are natural cheap communications transmitters, and CMOS image sensors are equipped in pervasive consumer electronics as detectors. New applications emerge with these tiny optical sensors. Image Sensor Communications (ISC) and positioning is easily realized based on CMOS sensors built upon smart phones. Millions of pixels can be maximally explored for communications~\cite{Huang:16} and accurate positioning~\cite{7636782}. The LED screen-to-camera communications constitute another interesting application, which may facilitate unimaginable potentials in entertainment, education, broadcasting and near-field communications, etc. Several early experiments have been already carried out with exciting findings, including projects of SoftLight~\cite{7524510}, Uber-in-light~\cite{7524513} and SBVLC~\cite{7061506} etc. In addition to traditional LEDs, emerging Organic LEDs (OLEDs) are attractive in flexibility, easy integration and fabrication, convenient color selectivity, and wide viewing angle. Thus they serve as wearable and portable VLC transmitters as well as fixed-location large screen communication transmitters~\cite{6617716,6809174,7833617}. Application domains encompass those of LEDs as well as new body area networks and sensor networks. Last but not least, LED based lighting infrastructures are becoming ubiquitous and have already been dubbed as the `eyes and ears of the IoT' in the context of smart lighting systems. Hence, VLC is a promising solution for indoor communications to a broad class of smart objects in particular in scenarios with high density of IoT devices. Steps along this direction have been presented in~\cite{Schmid,Li:2015:REB,7387378}.

\subsubsection{Commercialisation}
To create a larger economic and societal impact, the above mentioned academic research has to be in collaboration with industries. More importantly, the future success of VLC urges the joint efforts from both information industry and illumination industry, driven by demands from verticals, such as health, automotive, manufacturing, entertainment etc. The inclusive of mobile phones within the ecosystem would be highly desirable to the wide public acceptance. One promising area for real-world VLC to grow is the future pervasive IoT in consumer electronics, retailers, warehouses, offices and hospitals, etc. For example, Philips has recently commercialised VLC aided indoor localisation for the hypermarket in France~\footnote{\url{http://www.lighting.philips.co.uk/systems/themes/led-based-indoor-positioning.html}}. They have also teamed up with Cisco on an exciting IoT project called `Digital Ceiling'~\footnote{\url{http://www.cisco.com/c/en/us/solutions/digital-ceiling/partner-ecosystem.html}}, which connects all of a building's services in a single and converged network, where empowered LEDs can be used to collect, send and analyse data.

\section{Conclusion}
\label{sec_con}
This paper provides an overview on the localization, communications and networking aspects of VLC, along with the discussions on various challenges and opportunities. It is envisioned that the future research of VLC will open up new scientific areas in the wider academic context, which will be extremely beneficial to the whole community. VLC will create a unique opportunity to innovate all areas related to the future evolution, deployment and operation of ultra-dense small-cell networks, where potentially hundreds of people and thousands of appliances will be connected. More broadly, VLC will stimulate commercial solutions for supporting new killer applications, facilitating innovations for entertainment, collaborative design and vehicular networking, etc. By simultaneously exploiting illumination and communications, VLC will directly contribute towards the all-important `green' agenda, which has been one of the salient topics in the 21st-century.~\nocite{6283481,rong1,rong2,7542205,7506307,7437435,7096279,7010910,7180515,6877673,6644231,6685605,6470760,5640682,5692128,5522468,5378540,5641646,5337995,4663889,4804718,4813278,5161292,4451790,4460776,rong3,rong4,rong5,Wang2016,7802594,7542524,7465687,Wang:16,7533478,7552518,7247754,7217842,7468514,7482661,7061488,7217841,7056535,6933944,7124415,7008542,7018092,6670119,6670094,6476610,6363620,Li:13,Zhou:13,Zhou:12,6145718,6226483,6036194,4682680,5336786,Li,7249136,6399185,6214337,6213936,6093299,6093303,5779142,5779355,5779343,5594187,5594577,5594090,5594215,5502467,5493939,5208101,5073827,5073801,4698519,4657193,4533934,4525970,4489118,4224386}


\bibliographystyle{IEEEtran}
\bibliography{ref,rong_pub}

\end{document}